\documentclass[11pt,a4paper]{article}
 
\usepackage{graphicx}
\usepackage{epstopdf}
\usepackage{jcappub}

\usepackage{amsmath,amsthm,latexsym,amssymb,amsfonts,epsfig}
\usepackage{fontenc,dsfont,layout}
\usepackage{hyperref}
\usepackage{psfrag}
\usepackage[cp1250]{inputenc}

\usepackage[usenames,dvipsnames]{xcolor}

\numberwithin{equation}{section}

\title{Inflation and dark energy from the Brans-Dicke theory}
\author[a]{Micha{\l}~Artymowski}
\author[b]{Zygmunt~Lalak}
\author[c]{Marek~Lewicki}

\affiliation[a]{Institute of Physics, Jagiellonian University\\
{\L}ojasiewicza 11, 30-348 Krak{\'o}w, Poland}
\affiliation[b,c]{Institute of Theoretical Physics, Faculty of Physics, University of Warsaw\\ 
ul. Ho\.{z}a 69, 00-681 Warszawa, Poland}

\emailAdd{Michal.Artymowski@uj.edu.pl}
\emailAdd{Zygmunt.Lalak@fuw.edu.pl}
\emailAdd{Marek.Lewicki@fuw.edu.pl}

\abstract{We consider the Brans-Dicke theory motivated by the $f(R) = R + \alpha R^n -
\beta R^{2-n}$ model to obtain a stable minimum of the Einstein frame scalar
potential of the Brans-Dicke field. As a result we have obtained an inflationary scalar
potential with non-zero value of residual vacuum energy, which may be a source of
Dark Energy. In addition we discuss the probability of quantum tunnelling from the minimum of the potential. Our results can be easily consistent with PLANCK or BICEP2 data for appropriate choices of the value of $n$ and $\omega$.}


\begin{document}
\maketitle

\section{Introduction}

Inflationary models \cite{Lyth:1998xn,Liddle:2000dt,Mazumdar:2010sa,Starobinsky:1980te} are widely accepted as a solution to problems of classical cosmology, such as flatness, horizon and monopole problem. They may also be responsible for the generation of primordial inhomogeneities and their predictions on that are consistent with astronomical observations of cosmic microwave background (CMB) \cite{Ade:2014xna,Ade:2014gua,Spergel:2006hy,Ade:2013zuv}. During the inflationary era one obtains the accelerated growth of the FRW scale factor, which is ended by the so-called reheating characterized by the generation of relativistic particles. Inflation may be generated by e.g. scalar fields minimally or non-minimally coupled to gravity. The latter case will be analysed in this paper.
\\*

On the other hand the series of experiments \cite{Perlmutter:1999jt,Spergel:2006hy,Ade:2013zuv} convincingly suggests the existence of the so-called Dark Energy (DE) with barotropic parameter close to $-1$. One of the possible sources of DE may be  a non-zero vacuum energy of a scalar field, which in principle can be the $f(R)$ theory \cite{DeFelice:2010aj,Capozziello:2002rd,Copeland:2006wr} or the Brans-Dicke field. 
\\*

The Brans-Dicke theory is widely discussed in the context of cosmic inflation \cite{Starobinsky:1994mh}, dark energy (DE) \cite{Tsujikawa:2010zza}, stability of stars \cite{Babichev:2009fi},  gravitational collapse and quantum gravity corrections \cite{Artymowski:2013qua,Zhang:2012em}. It may be interpreted as a generalisation of $f(R)$ theory with an auxiliary field $\varphi : = F(R) = f'(R)$, where prime denotes the derivative with respect to the Ricci scalar $R$. The Brans-Dicke theory may be expressed in the Jordan frame (where the non-minimal coupling of $\varphi$ to the gravity is explicit) or in the Einstein frame (where the conformal transformation modifies the action into its canonical, minimally coupled form). In the classical theory (i.e. without quantum gravity corrections) those two approaches are equivalent to each other.
\\*

In \cite{Artymowski:2014gea} we have analysed the $R + \alpha R^n$ theory, which appeared to be a realistic inflationary model. Under some assumptions it has a stable minimum which provides the graceful exit and reheating of the universe. Depending on the value of the $n$ parameter its primordial inhomogeneities may be consistent with PLANCK \cite{Ade:2013zuv} or BICEP2 \cite{Ade:2014xna,Ade:2014gua} data. Similar analysis was also performed in the Ref. \cite{Motohashi:2014tra}. By extending this model into $f(R) = R + \alpha R^n - \beta R^{2-n}$ we have obtained an improved model, which provides a minimum of an Einstein frame scalar potential for all values of the $n$ parameter as well as n non-zero vacuum energy of order of  $(n\delta)^{1/(n-1)}$, which could be responsible for the present acceleration of the Universe.
\\*

{In this paper we want to extend this analysis into the Brans-Dicke theory by taking non-zero value of the $\omega_{BD}$. We keep scalar potentials in the same form they appeared in \cite{Artymowski:2014gea} and we look for analytical and numerical solutions for the evolution of the background as well as for the primordial perturbations.}
\\*

The structure of this paper is as follows. In the Sec. \ref{sec:Infl} we analyse the $U\propto (\varphi-1)^{n/(n-1)}$ model: we discuss analytical inflationary solutions and primordial inhomogeneities. In the Sec. \ref{sec:DE} we generalise this model to obtain the non-zero vacuum energy of the Einstein frame potential in its minimum. In the Sec. \ref{sec:DENumerical} we discuss numerical study of the evolution of the model with dust modelling the contribution of  matter fields to energy density. In the Sec. \ref{sec:tunnel} we investigate the {stability of the DE solution and the} possibility of quantum tunnelling of $\varphi$ from the minimum of the Einstein frame scalar potential. Finally, we conclude in the Sec. \ref{sec:concl}.

\section{The modified Starobinsky inflation} \label{sec:Infl}

Let us consider a {Brans-Dicke theory} in the flat FRW space-time with the metric tensor of the form {$ds^2 = -dt^2 + a(t)^2(d\vec{x})^2$}. Then the {Jordan frame} action is of the form
\begin{equation}
S =  \int d^4x \sqrt{|g|}\left[\varphi R - \frac{\omega}{2\varphi}(\nabla\varphi)^2 - U(\varphi)\right] + S_{\text{m}}\, , \label{eq:actionJ}
\end{equation}
where $\omega = const$ and $S_{\text{m}}$ is the action of matter fields. Then, for the homogeneous field $\varphi$ the field's equation of motion and the first Friedmann equation become \cite{DeFelice:2010aj}
\begin{eqnarray}
\ddot{\varphi} + 3H\dot{\varphi} + \frac{2}{\beta}(\varphi U_\varphi - 2U) &=& \frac{1}{\beta}\left(\rho_M - 3P_M\right)\ ,\label{eq:motionBD}\\
3\left(H + \frac{\dot{\varphi}}{2\varphi}\right)^2 &=& \frac{\beta}{4}\left(\frac{\dot{\varphi}}{\varphi}\right)^2 + \frac{U}{\varphi} + \frac{\rho_M}{\varphi}\, , \label{eq:FriedBD}\\
\dot{\rho}_M + 3H(\rho_M + P_M) &=& 0\, , \label{eq:cont}
\end{eqnarray}
where $\beta = 2\omega+3$, $U_\varphi:=\frac{dU}{d\varphi}$ and $\rho_M$ and $P_M$ are energy density and pressure of matter fields respectively \footnote{{In this paper we refer as matter fields to all perfect fluid components of the energy-stress tensor, like dust, radiation or scalar fields.}}. Let us note that $U$ may be interpreted as a energy density, but $U_\varphi$ is not an effective force in the Eq. (\ref{eq:motionBD}). One can define the effective potential and its derivative - the effective force, by
\begin{equation}
U_{eff} = \int \frac{2}{\beta}(\varphi U_\varphi-2U) d \varphi - \varphi  \frac{1}{\beta}(\rho_M-3p_M) + C\, , \qquad F_{eff} := \frac{d U_{eff}}{d\varphi}\, , 
\end{equation}
where $C$ is unknown constant of integration. The effective potential shall be interpreted as a source of an effective force, but not as a energy density.
\\*

Let us consider a generalisation of the Starobinsky inflation motivated by the $f(R)$ theory from the Ref. \cite{Codello:2014sua}, namely
\begin{equation}
U = \frac{n-1}{2}\alpha\left(\frac{\varphi-1}{n\alpha}\right)^\frac{n}{n-1} \, . \label{eq:Uinfl}
\end{equation}
In the $n=2$ case the Eq. (\ref{eq:Uinfl}) recovers the Starobinsky potential. As mentioned in the Ref. \cite{Artymowski:2014gea} the considered potential has a minimum {in the Einstein frame} only for certain values of $n$ parameter. For other values of $n$ it may obtain negative or complex values, which requires additional terms in the potential. Thus, in order to obtain the graceful exit and reheating for all $n$ one needs to extend this potential into more general form, which shall be done in further parts of this paper. Different modifications of the Starobinsky model were also discussed in Ref. \cite{Ben-Dayan:2014isa,Sebastiani:2013eqa}.

\subsection{Slow-roll solutions} \label{sec:analytical}

During the cosmic inflation the field is in its slow-roll regime, which means that $\ddot{\varphi}\ll3H\dot{\varphi}$ and $\dot{\varphi}\ll H\varphi$. In such a case, for the potential from the Eq. (\ref{eq:Uinfl}) the number of e-folds until the end of inflation is equal to
\begin{equation}
{N \simeq \frac{\beta}{2}\int\frac{U d\varphi}{\varphi(\varphi U_\varphi - 2U)} = \frac{\beta}{4}\left(\frac{n}{2-n}\log \left(2-\varphi+2\frac{\varphi-1}{n}\right)-\log(\varphi )\right) \, .}\label{eq:efolds}
\end{equation}
The initial conditions were chosen to satisfy $N(\varphi=1)=0$, since $\varphi=1$ is a typical value of the field at the end of inflation. This result is approximately the same for the Einstein frame analysis in the slow-roll regime. 
\\*

Let us assume that during the slow-roll era we are in the $\varphi\gg 1$ regime {and that at the moment of 50-60 before the end of inflation $\varphi$ satisfies $\varphi\,\beta\,(2-n)\gg(1-n)^2$}. Then the Jordan frame potential satisfies $U \propto \varphi^{n/(n-1)}$ and Eq. (\ref{eq:motionBD},\ref{eq:FriedBD}) have following slow-roll solutions
\begin{equation}
\varphi \simeq n\alpha\left(\sqrt{\frac{n-1}{6n}}\frac{2-n}{\beta(n-1)^2}t\right)^{-2(n-1)}\, , \qquad H\simeq \frac{1}{\epsilon t}:= \beta\frac{(n-1)^2}{(2-n)t} \, . \label{eq:SRBD}
\end{equation}
In such a case one obtains a power-law inflation with a constant slow-roll parameter $\epsilon$ and a scale factor proportional to $t^{1/\epsilon}$. Hubble parameter needs to be positive in order to obtain the expansion of the universe. Thus, from the Eq. (\ref{eq:SRBD}) one requires $n<2$. 
Inflationary evolution appears for $\epsilon<1$, which gives the lower limit for $n$. Finally one obtains the following allowed range for $n$
\begin{equation}
n \in \left(1+\frac{1}{2\beta}(\sqrt{1+4\beta}-1),2\right)
\end{equation}

\subsection{Einstein frame analysis}

\begin{figure}[h]
\centering
\includegraphics[height=5.8cm]{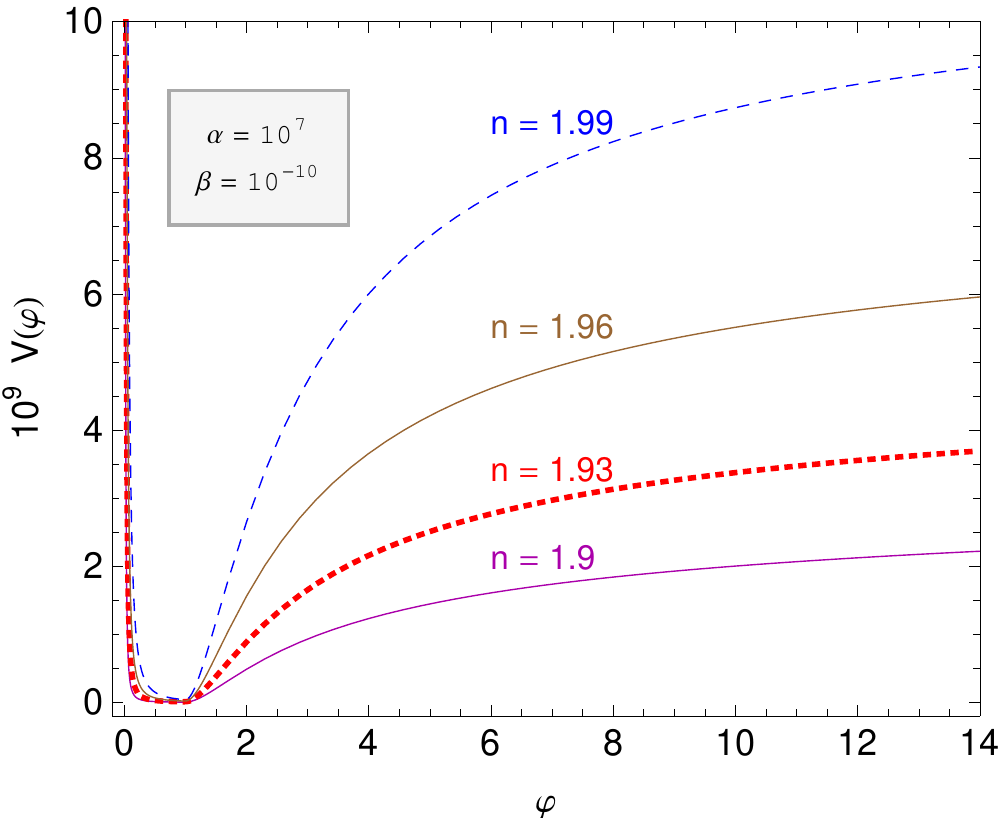}
\hspace{0.5cm}
\includegraphics[height=5.8cm]{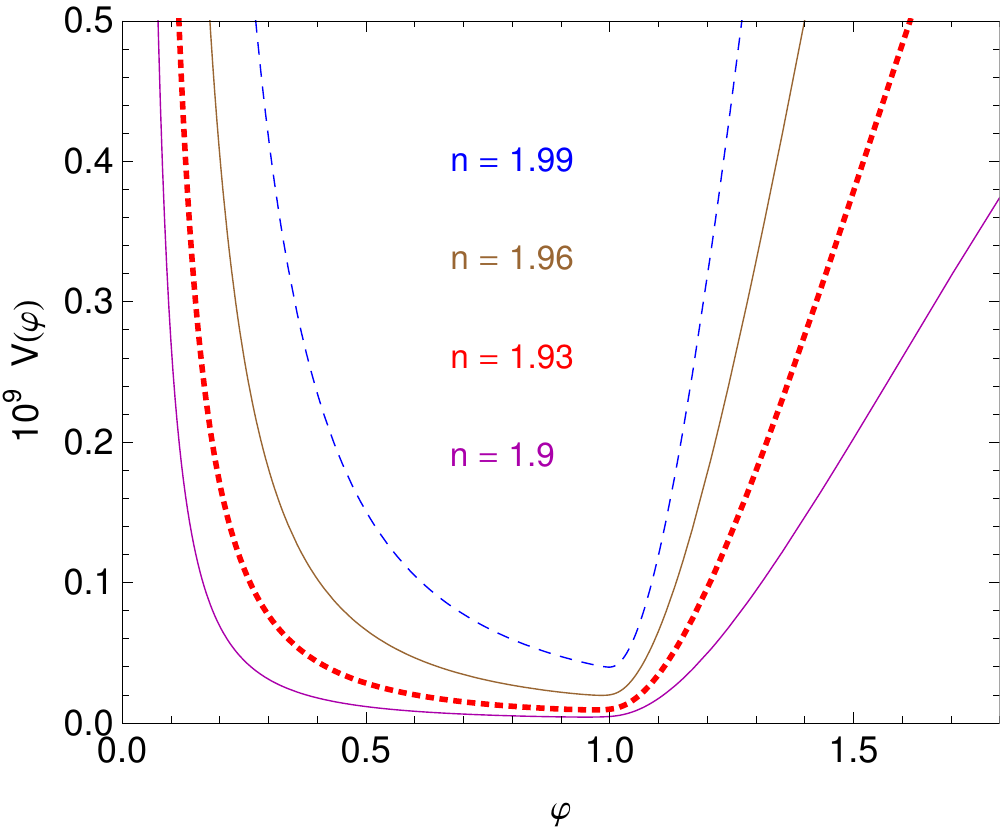}\\
\vspace{0.5cm}

\includegraphics[height=5.8cm]{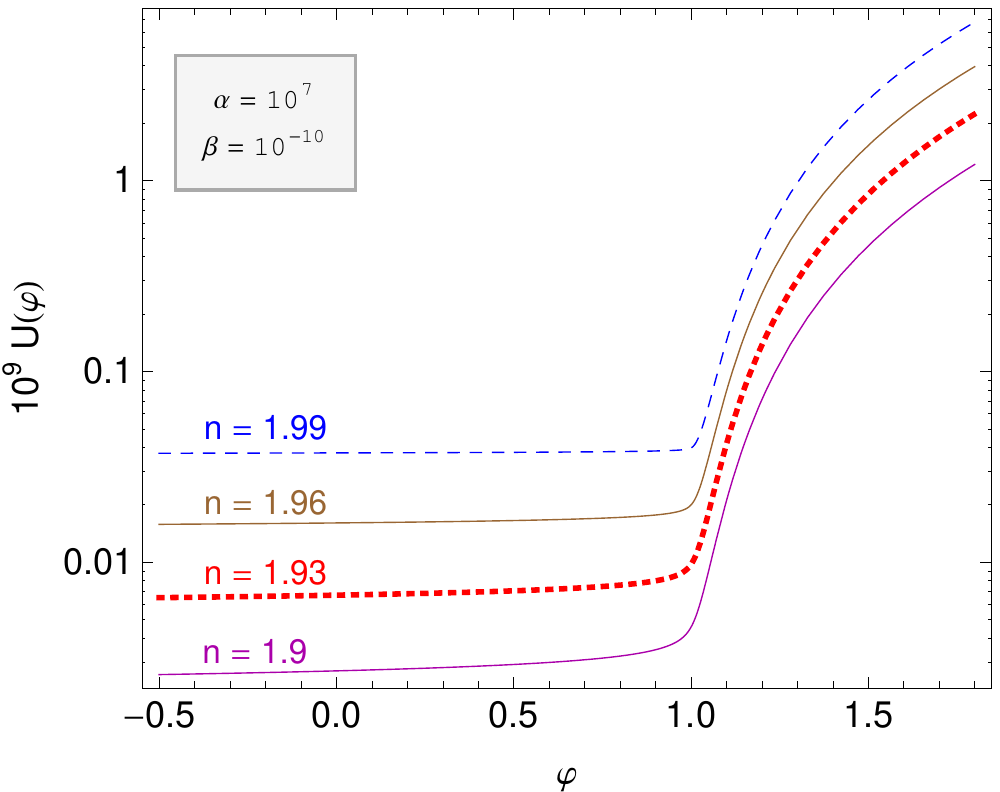}
\hspace{0.4cm}
\includegraphics[height=5.8cm]{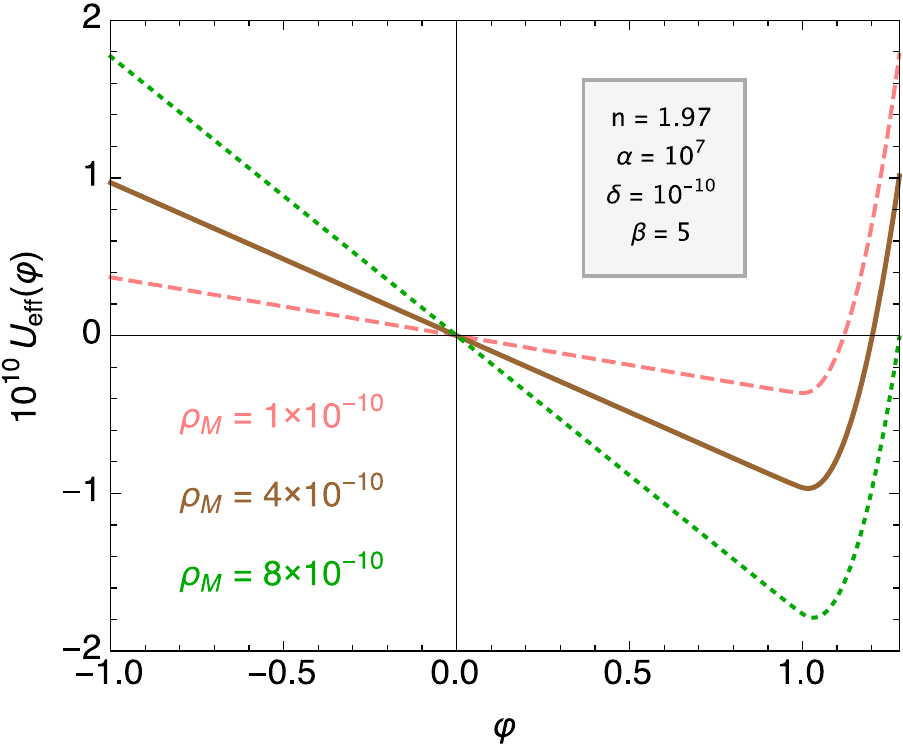}
\caption{\it Upper panels show the Einstein frame scalar potential from the Eq. (\ref{eq:DEpotential}) for different values of $n$. Inflation takes place for $\varphi\gg1$ and it ends when $\varphi$ reaches the local minimum of $V(\varphi)$ at $\varphi=\varphi_{min}\lesssim 1$. {The $\varphi<0$ regime} is separated from the minimum by the infinite wall of the potential at $\varphi=0$. Lower left panel presents the Jordan frame scalar potential $U(\varphi)$ for different $n$. The $U(\varphi)$ always decreases with $\varphi$ and it has no minimum. The lower right panel shows the effective potential $U_{eff}(\varphi)$ (which is the source of the effective force in the EOM) for different values of $\rho$. The minimum is deeper and it is getting closer to $\varphi = 1$ as the $\rho_M$ starts to dominate over $U(\varphi)$.}
\label{fig:Potential_m+1}
\end{figure}

The gravitational part of the action may obtain its canonical (minimally coupled to $\varphi$) form after transformation to {the} Einstein frame. {Let us assume that $\varphi>0$. Then for the Einstein frame metric tensor}
\begin{equation}
\tilde{g}_{\mu\nu}=\varphi g_{\mu\nu}\, , \qquad d\tilde{t}=\sqrt{\varphi}dt\, ,\qquad\tilde{a} = \sqrt{\varphi}a
\end{equation}
{one obtains} the action of the form of {
\begin{equation}
S[\tilde{g}_{\mu\nu},\varphi] = \int d^4\tilde{x} \sqrt{-\tilde{g}}\left[ \frac{1}{2}\tilde{R} - \frac{\beta}{4}\left( \frac{\tilde{\nabla}\varphi}{\varphi} \right)^2 - \frac{U(\varphi)}{\varphi^2} \right] + S_m[\tilde{g}_{\mu\nu},\varphi]\, ,
\end{equation}}
where $\tilde{\nabla}$ is the derivative with respect to {the Einstein frame coordinates. Matter fields are now explicitly coupled to $\varphi$ due to the fact that $d^4x\sqrt{-g} = \varphi^{-2}d^4\tilde{x}\sqrt{-\tilde{g}}$}. In order to obtain the canonical kinetic term for $\varphi$ let us use the Einstein frame scalar field $\phi$ 
\begin{equation}
\phi = \sqrt{\frac{\beta}{2}}\log\varphi \, , \qquad \varphi = \exp\left(\sqrt{\frac{2}{\beta}}\phi\right) \,  .
\end{equation}
The action in terms of $\tilde{g}_{\mu\nu}$ and $\phi$ looks as follows
\begin{equation}
S = \int d^4x \sqrt{-\tilde{g}}\left[ \frac{1}{2}\tilde{R} - \frac{1}{2}\left( \tilde{\nabla}\phi \right)^2 - V(\phi)\right]\, ,
\end{equation}
where
\begin{equation}
V = \left.\frac{U(\varphi)}{\varphi^2}\right|_{\varphi=\varphi(\phi)}
\end{equation}
is the Einstein frame scalar potential. Let us note that $(\varphi U_\varphi-2U)$ from the Eq. (\ref{eq:motionBD}) can be expressed as $\varphi^3V_\varphi$, so the minimum of $V$ shall also be the minimum of the effective potential in the Jordan frame. In fact, most important features of the potential, like existence of minima and barriers between them, which determine the evolution of the field in the Einstein frame are reflected in  the evolution of the field in the Jordan frame. In further parts of this paper we will often refer to the Einstein frame potential, even though we consider the Jordan frame as the primordial (or defining) one. Since all of the analysis performed in this paper is classical, descriptions in both frames give the same physical results. However, the description in the Einstein  frame is more intuitive, due to the canonical form of the scalar field's kinetic term and the minimal coupling between the field and the gravity. The only exception is the $\varphi\to 0$ limit, which usually leads to $V\to\infty$ due to the $\varphi^{-2}$ term in the potential. This infinity comes from the singularity of the Einstein frame metric tensor and it does not appear in the Jordan frame analysis, neither in $U$ nor in $U_{eff}$. The infinite barrier of the Einstein frame potential {comes from the singularity of $\tilde{g}_{\mu\nu}$ at $\varphi=0$ and therefore is non-physical}. There are several Brans-Dicke or $f(R)$ models of Dark Energy (see e.g. the Ref. \cite{Amendola:2006we}), where $\varphi$ may pass $\varphi = 0$ and obtain negative values. Thus it is best to work in the Jordan frame while considering the evolution around $\varphi=0$. The comparison between Jordan and Einstein frame potentials is shown at the Fig. \ref{fig:Potential_m+1}.
\\*

The Einstein frame scalar potential for $f(R) = R + \alpha R^n$ as a function of $\varphi$ has the following form
\begin{equation}
V(\varphi) = \alpha(n-1)(\alpha n)^{\frac{-n}{n-1}}\left(1-\frac{1}{\varphi}\right)^2(\varphi-1)^{\frac{2-n}{n-1}}\, ,\label{eq:VRn}
\end{equation}
where the last term parametrizes the deviation from the Starobinsky potential. Let us note that the slow-roll solutions for the Brans-Dicke field, Hubble parameter and the number of e-folds are the same in Einstein and Jordan frames. In the $\phi/\sqrt{\beta}\ll 1$ limit one finds $\varphi\simeq 1+ \phi\sqrt{2/\beta}$ and the Einstein frame potential takes the following form
\begin{equation}
V(\phi)\simeq \alpha(n-1)\left(\alpha n\sqrt{\frac{\beta}{2}}\right)^{\frac{-n}{n-1}}\phi^{\frac{n}{n-1}} \, .
\end{equation}
The $\phi$ field is minimally coupled to gravity, so the model evolves like in the $V(\phi) = \lambda \phi^{\frac{n}{n-1}}$ in the GR frame.
\\*

{For $\varphi<0$ one needs to define the Einstein frame according to the following procedure: a) Define the Einstein frame metric tensor $\tilde{g}_{\mu\nu} = -\varphi g_{\mu\nu}$. Then $d\tilde{t} = \sqrt{-\varphi}dt$, $\tilde{a} = \sqrt{-\varphi}a$, where $\sqrt{-\varphi}$ remains real. b) Rewrite the action from the Eq. (\ref{eq:actionJ}) using $\tilde{g}_{\mu\nu}$ which gives the canonical for of the GR action with a negative sign and the BD field part with a ghost-like kinetic term. }

\subsection{The generation of primordial inhomogeneities}

\begin{figure}[h]
\centering
\includegraphics[height=5.7cm]{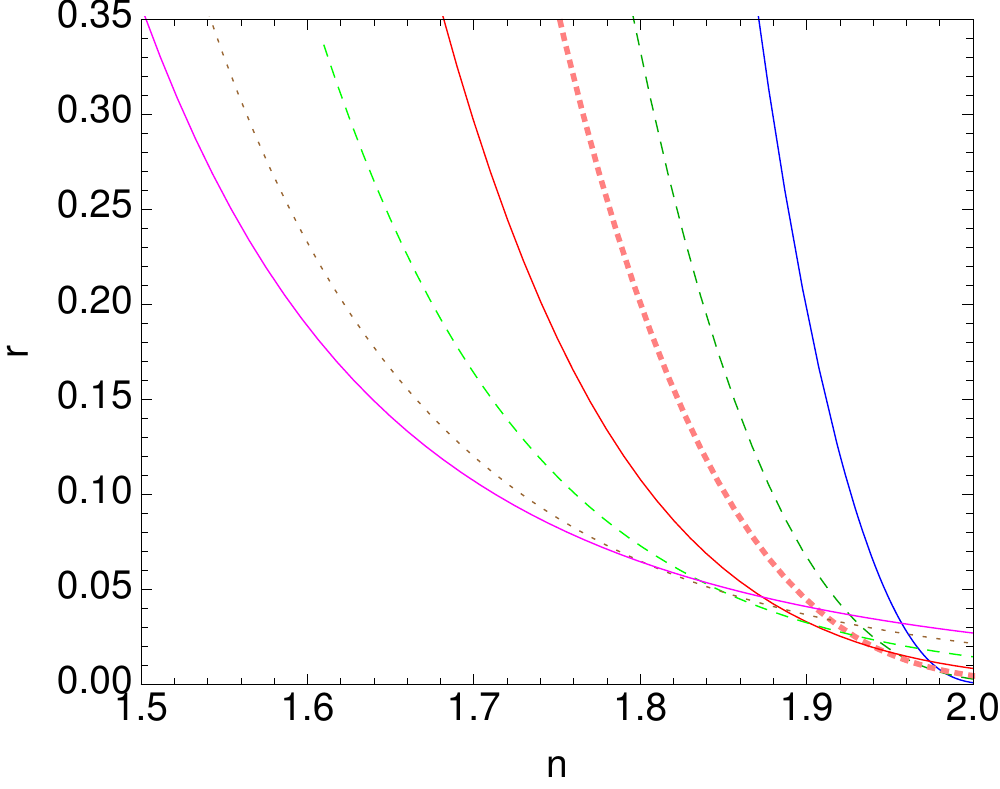}
\hspace{0.5cm}
\includegraphics[height=5.7cm]{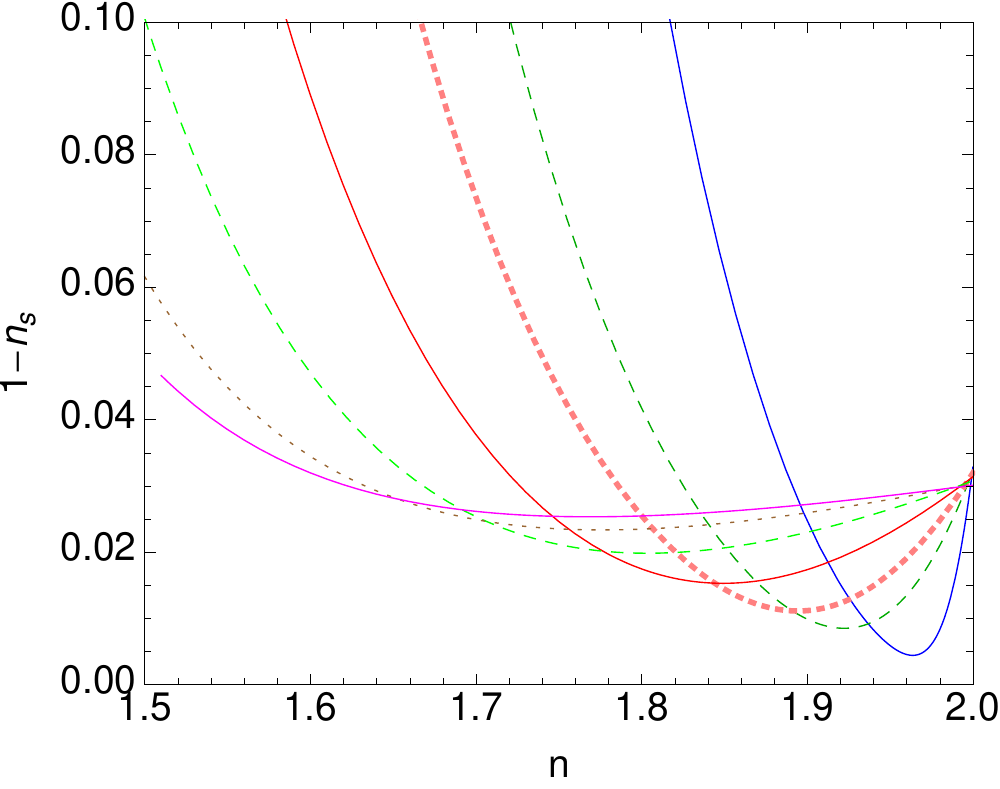}\\
\vspace{0.5cm}

\includegraphics[height=5.7cm]{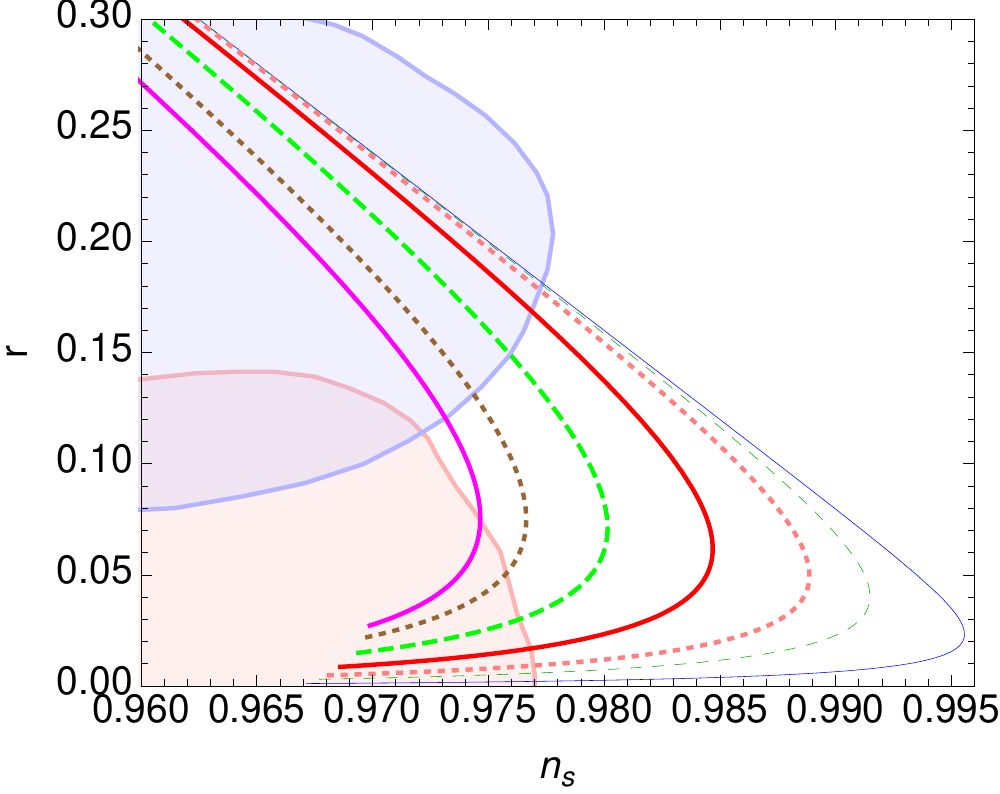}
\hspace{0.5cm}
\includegraphics[height=5.7cm]{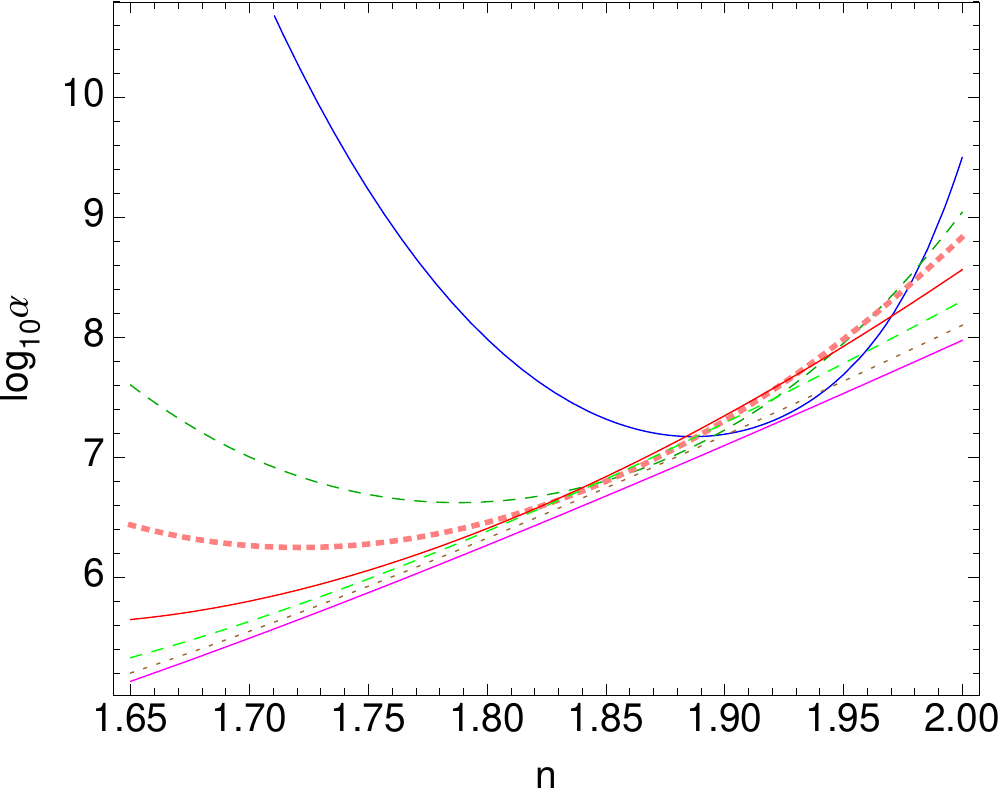}
\caption{\it Upper panels: The analytical solution for $r$ and $n_s$ as a function of $n$ and $\beta$ at the moment of $N=60$ for $\beta = 1$ (blue solid line), $\beta = 3$ (green dashed line line), $\beta = 5$ (dotted orange line), $\beta = 10$ (red line), $\beta = 20$ (light green dashed line), $\beta = 35$ (brown dotted line) and $\beta = 50$ (pink line). For big values of $\beta$ the $n$ dependence of $n_s$ is very weak. Lower left panel: The $(r,n_s)$ plane for different $n$ and $\beta$. Straight lines, which appear for smaller values of $n$ (i.e. bigger values of $r$) represent the analytical solution from the Sec. \ref{sec:analytical}. {Blue and Red regions represent $2\sigma$ regimes of BICEP2 and PLANCK data respectively. Lower right panel:} The analytical solution for $\alpha$ as a function of $n$ at the moment of  $N=60$ calculated from the normalisation of perturbations. Note that for big $\beta$ the $\beta$-dependence of $\alpha$ is weaker.}
\label{fig:perturbations}
\end{figure}

The simplest way to calculate the power spectrum of primordial inhomogeneities is to quantise curvature perturbations in the Einstein frame. As shown in the Ref. \cite{DeFelice:2010aj} this procedure performed in the slow-roll regime gives the same results as the Jordan frame quantisation. Then one obtains
\begin{equation}
\mathcal{P}_{\tilde{\mathcal{R}}} = \left(\frac{\tilde{H}}{2\pi}\right)^2\left(\frac{\tilde{H}}{\phi'}\right)^2\, , \label{eq:PowerSpectrumE}
\end{equation}
where
\begin{equation}
\phi':=\frac{d\phi}{d\tilde{t}}\qquad  \text{and}\qquad  \tilde{H}:=\frac{1}{\tilde{a}}\frac{d\tilde{a}}{d\tilde{t}} = \frac{\tilde{a}'}{\tilde{a}}\, .
\end{equation}
In terms of the Jordan frame variables the Eq. (\ref{eq:PowerSpectrumE}) can be expressed as
\begin{equation}
\mathcal{P}_{\tilde{\mathcal{R}}} \simeq \mathcal{P}_{\mathcal{R}} = \frac{\beta}{24\pi^2}\frac{U^3}{\varphi^2\left(\varphi  U_{\varphi }-2U\right)^2} \, . \label{eq:PowerSpectrumJ}
\end{equation}
The spectral index $n_s$ and the tensor-to-scalar ratio $r$ are approximately equal to
\begin{equation}
n_s - 1 \simeq \frac{2}{U^2 \beta }\left(\varphi  \left(6 U U_{\varphi }-3 \varphi  U_{\varphi }^2+2 \varphi  U U_{\varphi \varphi }\right)-4 U^2\right)\, ,\quad r \simeq \frac{16}{\beta}\left(\frac{\varphi U_{\varphi}-2U}{U}\right)^2 \, .
\end{equation}
On the lower left panel of the Fig. \ref{fig:perturbations} we present the power spectrum on the $(n_s,r)$ plane for the inflationary potential (\ref{eq:Uinfl}) as the function of $n$ and $\beta$.

The normalisation of primordial inhomogeneities requires that $\mathcal{P}_\mathcal{R}^{1/2}\sim 5\times 10^{-5}$ at the moment of $50$ to $60$ e-folds before the end of inflation. One can use the normalisation of the power spectrum to obtain $\alpha$ as a function of $n$ and $\beta$. The result obtained from Eq. (\ref{eq:Uinfl},\ref{eq:efolds},\ref{eq:PowerSpectrumJ}) in the slow-roll regime is plotted at the Fig. \ref{fig:perturbations}.
\\*

The issue of primordial non-Gaussianities in the Brans-Dicke theory was widely analysed in the Ref. \cite{DeFelice:2011zh}. The authors argue that $|f_{NL}^{local}|\ll1$, since in the slow-roll regime of the single field inflation the $f_{NL}^{local}$ is proportional to slow-roll parameters. On the other hand the $f_{NL}^{equil}$ (related to non-standard kinetic terms) is {in the Brans-Dicke theory} equal to 
\begin{equation}
f_{NL}^{equil}\simeq -\frac{5}{4}\frac{\dot{\varphi}}{H\varphi}+\frac{5}{6}\frac{\ddot{\varphi}}{H\dot{\varphi}} \qquad\Rightarrow\qquad |f_{NL}^{local}| \ll 1\, .
\end{equation}
Thus our model does not produce significant amount of non-Gaussianities, which is consistent with the experimental data \cite{Ade:2013ydc}.

\begin{figure}
\centering
\includegraphics[height=5.7cm]{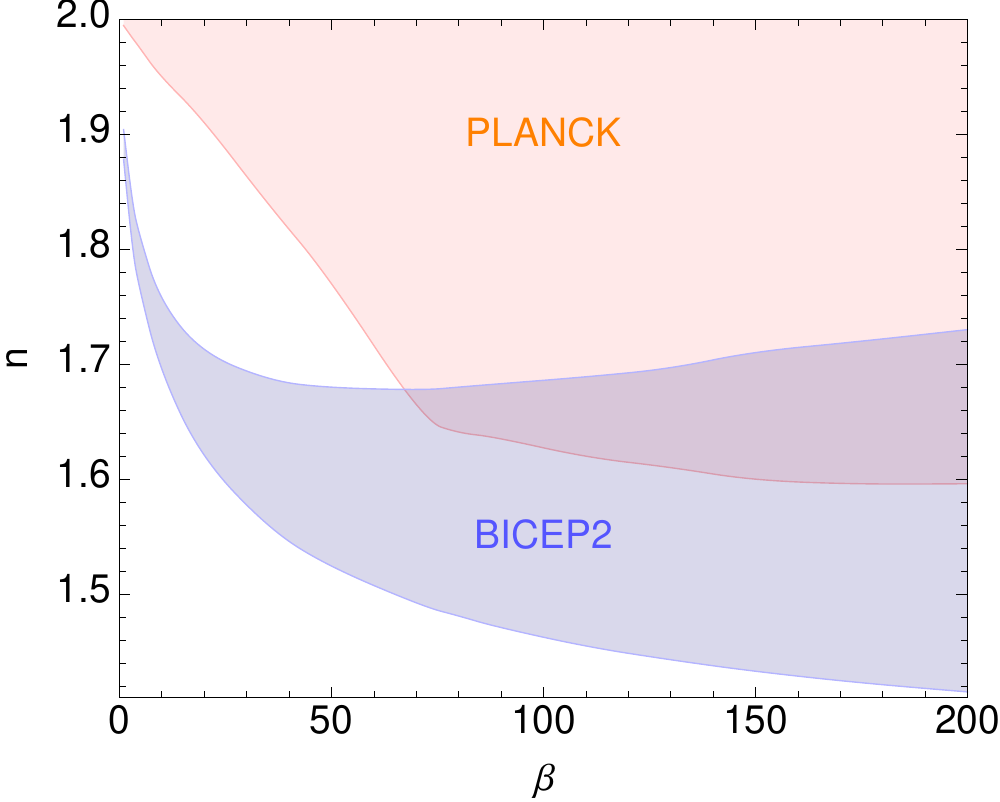}
\caption{\it {Regions on the $(\beta,n)$ plane for which the primordial inhomogeneities obtained for the potential (\ref{eq:Uinfl}) fit the Planck (red region) or BICEP2 (blue region) data. For $\beta\gtrsim 70$ and $n\sim 1.65$ one can fit the data of both experiments.}}
\end{figure}

\section{The dark energy model} \label{sec:DE}

Deficiencies of the model discussed in the previous Section can be bypassed by considering modification of the scalar potential motivated by the $f(R) = R + \alpha R^{n} - \delta R^{2-n}$ theory \cite{Artymowski:2014gea}, namely.  
\begin{equation}
U(\varphi) = \frac{1}{2} (n-1)\left(\alpha  \mathbb{R}^{ n}(\varphi )+\delta\mathbb{R}^{2-n}(\varphi )\right) \, ,\quad  V = \frac{1}{\varphi^2}U \, ,\label{eq:DEpotential}
\end{equation}
where
\begin{equation}
\mathbb{R}(\varphi) = \left(\frac{\sqrt{4 (2-n)n \alpha  \delta +(\varphi -1)^2}+\varphi -1}{2n \alpha }\right)^{\frac{1}{n-1}} \, .
\end{equation}
The $\mathbb{R}$ function could be interpreted as the Ricci scalar in the $f(R)$ theory. However in considered model {(for $\beta\neq 3$)} it has no connection with the curvature. Let us require $\alpha\gg1$,  $\delta\ll1$ and $\alpha\delta\ll1$. Then in the $\delta \to 0$ limit one restores the inflationary potential from the Section \ref{sec:Infl}. Both models give the same results under the slow-roll approximation and they generate the same power spectrum of primordial inhomogeneities during inflation. However, the potential from the Eq. (\ref{eq:DEpotential}) has several advantages: it has a minimum at $\varphi\sim 1$ for all values of $n$ in which it has a non-zero vacuum energy. In order to satisfy experimental data one needs $\delta\lll1$. Let us note that the DE solution lies in the stable minimum of the Einstein frame potential. This prevents the field from rolling down to the negative values of $\varphi$ at small energies, which is the case in e.g. Ref. \cite{Amendola:2006we}. Different model which combines Inflation and DE was described in the Ref. \cite{Nojiri:2003ft}
\\*

The Einstein frame potential has a minimum at  $\varphi_{min} \simeq \frac{2}{n} (n-1) (1+2 n \alpha  \delta )$. The minimum is slightly shifted with respect to  $\varphi=1$, which is the GR vacuum case. The value of $V$ at the minimum for small values of $\delta$ reads
\begin{equation}
V(\varphi_{min})\simeq \frac{n}{8 (n-1)^2} (n \delta )^{\frac{1}{n-1}} \left(n-1-n^2 \alpha  \delta \right) \sim \frac{1}{2}\delta^{\frac{1}{n-1}} \, .\label{eq:Vmin}
\end{equation}
Hence, this model predicts some amount of vacuum energy. 
\\*

Let us clarify that $\varphi_{min}$ {could be considered as} a local minimum of $V$. The $\mathbb{R}$ function (and therefore the potential $U$) has no global minimum, its minimal value is equal to 0 (at the $\varphi\to-\infty$ limit) and it continuously grows with $\varphi$. An example of Jordan and Einstein frame potentials and minima of $V(\varphi)$ are plotted at the Fig. \ref{fig:Potential_m+1}. The existence of  a stable minimum is one of the main differences between this model and DE models motivated by  $f(R) = R-\delta R^{2-n}$ theory, in which the auxiliary field {may roll} down towards negative $\varphi$ \cite{Amendola:2006we}. {As we will show} the minimum of the Einstein frame potential (visible at the Fig. \ref{fig:Potential_m+1}) prevents the $\varphi$ from obtaining negative values for any initial conditions with $\varphi(0)>0$ and real values of the Hubble parameter.

\subsection{Viability of the dark energy model}

Let us denote the present time as $t_0$. In order to obtain successful model of dark energy one needs to satisfy several conditions:
\begin{description}
\item 1) The kinetic term of the $\varphi$ is proportional to $\varphi^{-1}$. Thus, to avoid the ghost state {for $\beta>0$} one needs $\varphi_0:=\varphi(t_0)>0$. This condition is satisfied as long as the field lies in the minimum of its potential after inflation. The possibility of {passing $\varphi=0$ via the classical evolution or} quantum tunnelling is described in the Sec. \ref{sec:tunnel}
\item 2) To avoid the negative mass square for a scalar field degree of freedom one needs
\begin{equation}
U_{\varphi\varphi}>0 \qquad \text{for}\qquad t\leq t_0\, ,
\end{equation}
which means that $\mathbb{R}_0>0$. This condition is satisfied {for any values of $\varphi$}
\item 3) The model needs to obtain the correct low-energy limit and to satisfy consistency with local gravity constraints. This means that the low-energy action shall be of the form of 
\begin{equation}
S \simeq \int d^4x \sqrt{-g} \left(\frac{1}{2} R - \Lambda \right) + S_{m} \, ,
\end{equation}
After the $\varphi$ field is stabilised in its minimum (which in the dust domination era is exactly GR minimum) it produces the vacuum energy, which is a source of $\Lambda$. The $S_{m}$ comes from radiation and dust produced during the reheating of the universe.
l satisfies conditions for a viable DE model.
\end{description}

\section{Numerical analysis of the dark energy model} \label{sec:DENumerical}

\begin{figure}[h]
\centering
\includegraphics[height=4.6cm]{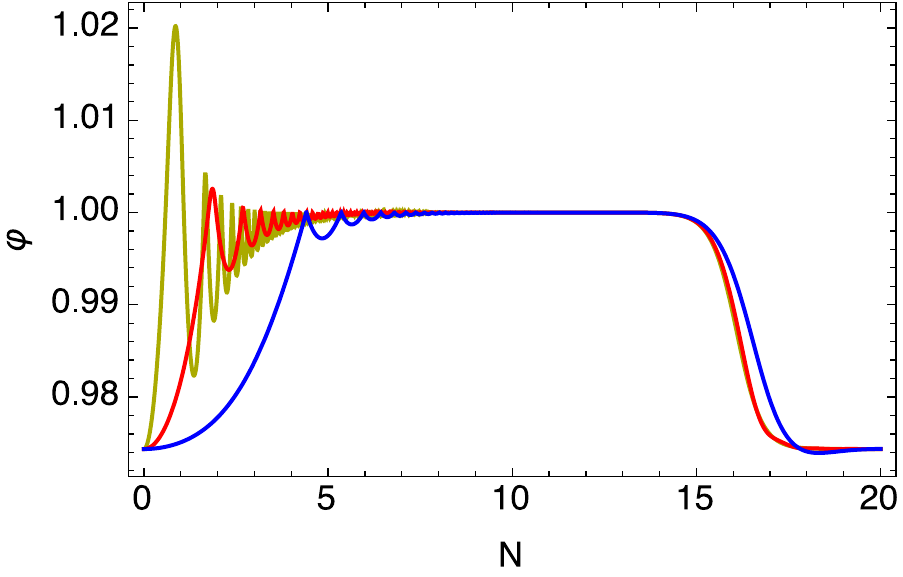}
\hspace{0.5cm}
\includegraphics[height=4.6cm]{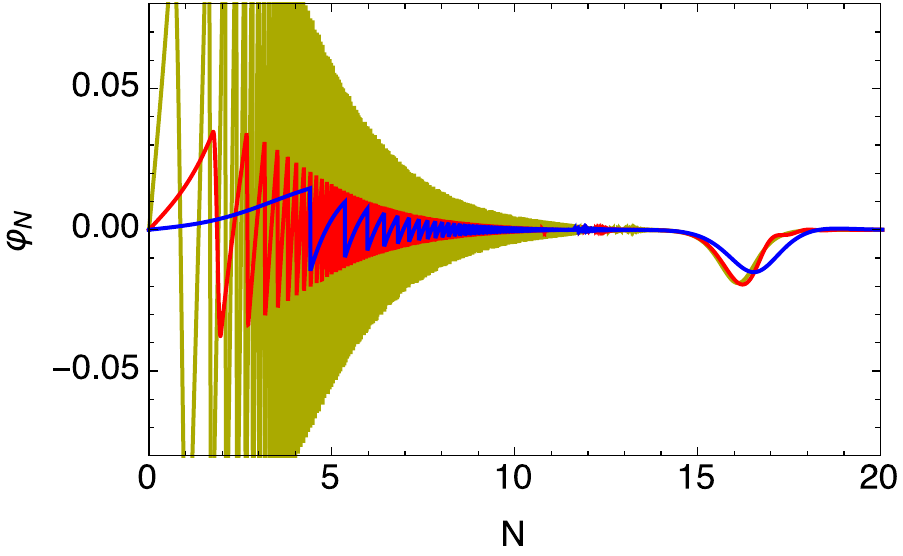}
\caption{\it Numerical results of the evolution of the Brans-Dicke field $\varphi$ and its derivative with respect to $N$ as a function of $N$. We have assumed $n=1.95$, $\alpha = 2\times 10^8$, $\delta = 10^{-30}$, $\varphi(0) = 1$, $\varphi_N(0) = 0$ {and , $\beta \in \{1,10,100\}$ (yellow, red and blue lines respectively)}. The $\rho_M$ {consists of} dust {and radiation} with $\rho_D(0)= 5\times10^{-11}$ {and $\rho_R(0) = 10^{-9}$}. {The $\varphi$ starts its evolution from $\varphi_{\min}$ (the radiation domination era) and rolls up towards $\varphi=1$ (dust domination era). This transition is more rapid for small values of $\beta$. When DE starts to dominate the field rolls down back to $\varphi_{\min}$.} The choice of initial conditions and parameters of the model is rather unrealistic (e.g. too big $\delta$), but it illustrates the way of the field to its minimum. The result is very similar to the one obtained in the $f(R)$ theory.}
\label{fig:phiDE}
\end{figure}

\begin{figure}[h]
\centering
\includegraphics[height=4.6cm]{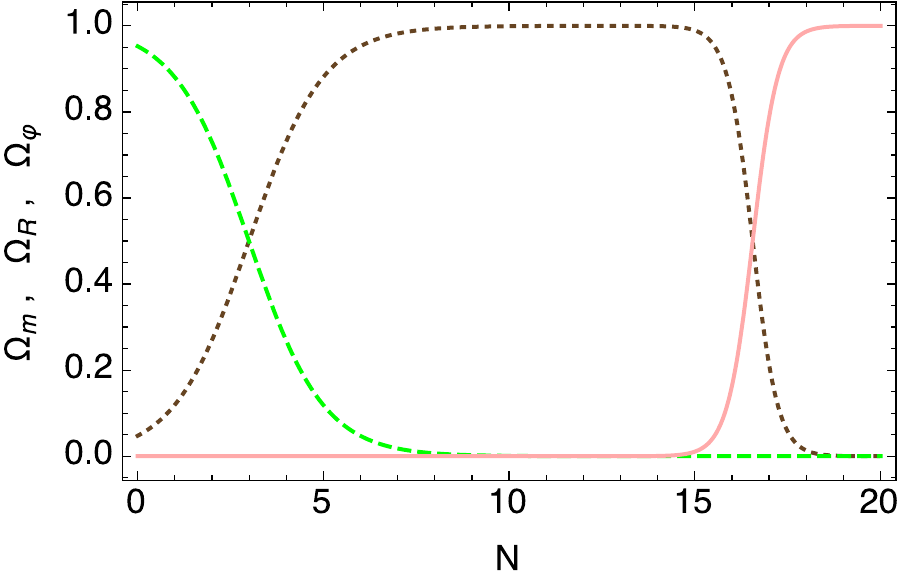}
\hspace{0.5cm}
\includegraphics[height=4.6cm]{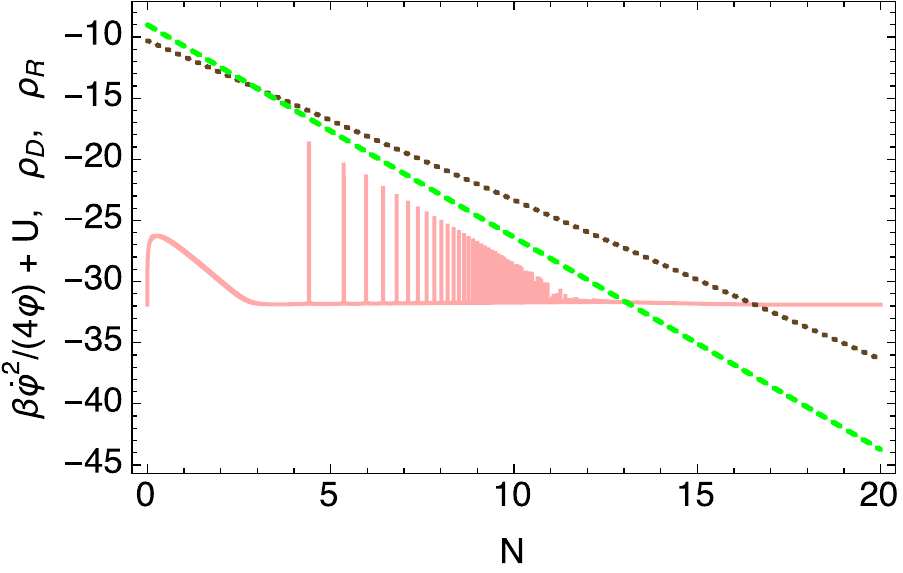}
\caption{\it Left panel: numerical result of the evolution of {density parameters $\Omega_m$, $
\Omega_R$ and $\Omega_\varphi$ (dotted brown, dashed green and pink lines respectively) as a function of N for $n=1.95$, $\alpha = 2\times 10^8$, $\delta = 10^{-30}$ and $\beta = 10$}. Note that the vacuum energy of the Brans-Dicke field starts to dominate when the field reaches {$\varphi_{\min}$}. Right panel: numerical result for the evolution of {energy density of dust, radiation and $\varphi$ (dotted brown, dashed green and pink lines respectively)}. The energy density of the $\varphi$ obtains the {equation of state} $\omega_\varphi\simeq-1$  about 10 e-folds before the DE domination period.}
\label{fig:HDE}
\end{figure}

The non-zero value of the Einstein frame potential at the minimum rises a possibility of obtaining a realistic solution to the dark energy problem. To analyse low energy solutions of the Jordan frame equations of motion let us use the number of e-folds (defined by $N:=\log(a)$) as a time variable. Then Eq. (\ref{eq:motionBD},\ref{eq:FriedBD}) read 
\begin{eqnarray}
H^2(\varphi_{NN}+3\varphi_N) + H_NH\varphi_N+\frac{2}{\beta}(\varphi U_\varphi-2U) &=& \frac{1}{\beta}(\rho_M-3p_M)\, ,\label{eq:NmotionBD}\\
H^2 &=& \frac{\rho_M + U}{3\varphi + 3\varphi_N + \frac{3-\beta}{4}\frac{\varphi_N^2}{\varphi}} \label{eq:NFriedBD}\, ,
\end{eqnarray}
where the index ``$ _N$'' denotes the derivative with respect to $N$. Since in the Jordan frame the Eq. (\ref{eq:cont}) is satisfied one finds $\rho_M=\rho_I e^{-3(1+w)N}$, where $w=p_M/\rho_M$ is a barotropic parameter. After the inflation $\varphi$ oscillates around $\varphi_{\min}$ and reheats the universe by the particle production. Thus, after oscillations one obtains the radiation domination era, for which $w=1/3$ and $\rho_M - 3p_M = 0$. The radiation increases the cosmic friction term but does not contribute to the $U_{eff}$, so the field is not shifted from the minimum of the potential $V$. However, during the dust domination era the $U_{eff}$ is modified and $\varphi$ oscillates around $\varphi=1$. The evolution of $\varphi$ and $\varphi_N$ during {radiation, dust and DE} domination eras is presented at the Fig. \ref{fig:phiDE}. The evolution of the Hubble parameter and $\sqrt{\rho_M/3}$ is plotted at the Fig. \ref{fig:HDE}. We have assumed that {the Universe is initially dominated by radiation and that} the field starts from the {$\varphi=\varphi_{\min}$. When the dust starts to dominate the field rolls up to $\varphi=1$, which is the GR limit of the theory.} When the dust becomes subdominant the $\varphi$ rolls to $\varphi_{min}$ and one obtains the Dark Energy with the barotropic parameter $\omega=-1$.
\\*

The evolution of $\varphi$, $\varphi_N$, $H$ and energy densities of $\varphi$ and dust in the $f(R)$ case (i.e. for $\beta=3$) have been presented at Fig. $8$ and $9$ in the Ref. \cite{Artymowski:2014gea}. As shown in the Fig. \ref{fig:HDE} {the energy density of $\varphi$} obtains the constant value when {$\varphi=\varphi_{\min}$. During that period one finds} $H, \varphi = const$, {which} implies that 
\begin{equation}
3H^2 \to \varphi_{min} V(\varphi_{min}) = 3\Omega_{DE} H_0^2 = const \, ,\label{eq:OmegaDE}
\end{equation}
where $\Omega_{DE}$ is a density parameter of DE. This equation is valid when DE completely dominates over the dust. From the Eq. (\ref{eq:OmegaDE}) one finds the connection between theoretical predictions of this model and astronomical observations.

{
\section{Stability of the vacuum} \label{sec:tunnel}
As discussed in previous sections, classical evolution of the field $\varphi$ brings it to rest at the minimum of its effective potential near $\varphi \approx 1$. However the Jordan frame potential $U(\varphi)$ (shown in the middle row of {the} Figure~\ref{fig:Potential_m+1}) {which gives the energy density}, 
has no minimum and decreases towards smaller values of field $\varphi$. This indicates that the vacuum at $\varphi\approx 1$ may actually be a metastable minimum and the \textit{true} vacuum may be located in the region of negative values of $\varphi$. In this Section we shall check if one can find initial conditions that could result in classical evolution ending at negative values of $\varphi$. We will also discuss {whether} quantum tunneling can allow our \textit{false} vacuum to decay into {a}  deeper global minimum.  
}

{
First we will show that classically the field can cross $\varphi=0$ {only if it starts from} unphysical initial conditions. In our model the field $\varphi$ can evolve towards negative values, {instead of rolling to} the minimum of the effective force, only if {the} Hubble parameter \eqref{eq:NFriedBD} is imaginary.
This happens because {the} imaginary Hubble parameter changes the sign of the kinetic part of scalar field equation of motion (\ref{eq:NmotionBD}), 
and the effective force (shown in lower right panel of Figure~\ref{fig:Potential_m+1}) pushes the field away from $\varphi=1$. In our model {the numerator} of the squared Hubble parameter \eqref{eq:NFriedBD} is always positive. Thus during evolution the sign of $H^2$ can only be changed by its denominator which would correspond to a point of infinite friction for the field. Hence, obtaining a ghost field is possible only if we set it as such from the beginning of its evolution.
}

{
To discuss quantum tunneling we will use the standard formalism \cite{Coleman:1977py,Coleman:1977kc}, assuming that vacuum decay proceeds through nucleation of true vacuum bubbles within our false vacuum.
Such bubble is an $O(4)$ symmetric scalar field configuration $\varphi=\varphi(\tau)$, with the metric given by $ds^2=d\tau^2 + r(\tau)^2(d\Omega)^2$. Here $d\Omega$ is an infinitesimal element of the $3D$ sphere, and $r(\tau)$ is the {radius} of that sphere. The resulting metric tensor is of the form of the $FRW$ metric with the curvature parameter $k=+1$. 
}
{ 
Euclidean action in Jordan Frame takes the form}
{\begin{equation}
S_E=2\pi^2 \int d\tau r^3 \left( \frac{\omega}{2 \varphi} \varphi_{\tau}^2  +U-\varphi R \right) + S_m[g_{\mu\nu,\ldots}]\, ,
\end{equation}
where $R=6\left( \frac{r_{\tau\tau}}{r}+\left( \frac{r_\tau}{r} \right)^2-\frac{1}{r^2} \right)$.} {The scalar field equation of motion reads}
{
\begin{equation}
\varphi_{\tau\tau}+3\frac{r_\tau}{r}\varphi_\tau=\frac{2}{\beta}(\varphi U_{\varphi}-2U) - \frac{1}{\beta}\left(\rho_M - 3P_M\right)\, .
\end{equation}
}
{
The first Friedman equation reads
\begin{equation}
3\left(\frac{r_\tau}{r} + \frac{\varphi_\tau}{2\varphi}\right)^2-\frac{1}{r^2} = \frac{\beta}{4}\left(\frac{\varphi_\tau}{\varphi}\right)^2-\frac{U}{\varphi} - \frac{\rho_M}{\varphi}\, ,
\end{equation}
and differs from \eqref{eq:motionBD}\eqref{eq:FriedBD} only by the term $1/r^2$ corresponding to the curvature of the sphere, and {by the} different sign of the effective force.
The appropriate boundary condition at the true vacuum (negative $\varphi$) is \cite{Guth:1982pn}:
\begin{eqnarray}
r=0, \\
\varphi_r=0 \,.
\end{eqnarray}
When the field evolves from such initial conditions in {inverted} effective potential from {the} Figure~\ref{fig:Potential_m+1}, it is clear the field will simply roll toward smaller values. And so no tunneling towards positive $\varphi$ is possible.
}


\section{Conclusions} \label{sec:concl}

In this paper we have analysed the Brans-Dicke theory motivated by $f(R) = R + \alpha R^n - \delta R^{2-n}$ model (with $\alpha\gg1$ and $\delta\ll1$) and we have obtained both features of the observable Universe in a single framework: we show how to obtain successful inflation and the non-zero residual value of the Ricci scalar in an extension of the BD Starobinsky-like model. In the Sec. \ref{sec:Infl} we have showed that the high energy limit of our theory, namely the Brans-Dicke theory with a Jordan frame scalar potential proportional to $(\varphi-1)^{n/(n-1)}$ can be easily consistent with PLANCK or BICEP2 data for appropriate choices of the value of $n$ and $\beta$. 
\\*

In the section \ref{sec:DE} we have considered a full theory, with a potential from the Eq. (\ref{eq:DEpotential}). In this case the Einstein frame scalar potential is real for all $\varphi$ and it has a minimum for all $n$. The potential has non-zero value at the minimum, which may become a source of DE. The value of the
 parameter $\alpha$ is set by the normalisation of primordial inhomogeneities, while the value of the parameter $\delta$ (as a function of $n$ and $\beta$) can be read from the measured value of the present DE energy density.
\\*

In the section \ref{sec:DENumerical} we have performed numerical analysis of the late-time evolution of the model with dust employed as a matter field. During the radiation domination era the $\rho_M-3p_M = 0$, so the effective potential in the Jordan frame obtains its vacuum form. Thus the field holds $\varphi=\varphi_{min}$. During the dust domination era one finds $\varphi = 1$, which corresponds to the GR limit of the theory. When matter starts to be subdominant the $\varphi$ rolls to its minimum in $\varphi=\varphi_{min}\lesssim1$. Even before that moment the energy density of the scalar field becomes constant and the energy density of the $\varphi$ evolves like DE with barotropic parameter $\omega_\varphi=-1$. 

Finally, we have considered the possibility of quantum tunnelling of the Brans-Dicke field to the region of its negative values. Inspection of the equations of motion in the presence of gravitational background leads to the conclusion, that there are no bounce solutions interpolating between positive and negative values of the BD field. On the other hand there exist rolling solutions connecting the two regions. However, they correspond to unphysical initial conditions. 
Hence one is lead to the conclusion, that the vacuum near $\varphi = 1$ is stable. 

\acknowledgments

\noindent The work of Z.L. and M.L. has been partially supported by the Polish NCN grant DEC-2012/04/A/ST2/00099. The work of MA has been supported by the Polish NCN grant FUGA UMO-2014/12/S/ST2/00243.

\end{document}